\documentclass[english,aps,preprint]{revtex4}
\usepackage[T1]{fontenc}
\usepackage[latin9]{inputenc}
\usepackage{graphicx}

\makeatletter

\providecommand{\tabularnewline}{\\}

\@ifundefined{textcolor}{}
{%
 \definecolor{BLACK}{gray}{0}
 \definecolor{WHITE}{gray}{1}
 \definecolor{RED}{rgb}{1,0,0}
 \definecolor{GREEN}{rgb}{0,1,0}
 \definecolor{BLUE}{rgb}{0,0,1}
 \definecolor{CYAN}{cmyk}{1,0,0,0}
 \definecolor{MAGENTA}{cmyk}{0,1,0,0}
 \definecolor{YELLOW}{cmyk}{0,0,1,0}
}

\makeatother

\makeatother

\usepackage{babel}

\makeatother

\usepackage{babel}
\begin{document}

\title{Single production of fourth family sneutrino via RPV couplings at
linear colliders}

\author{O. \c{C}ak\i{}r}

\email{ocakir@science.ankara.edu.tr}

\author{S. Kuday}

\email{kuday@science.ankara.edu.tr}

\affiliation{Department of Physics, Ankara University, Faculty of Sciences, Ankara,
Turkey}

\author{\.{I}.T. \c{C}ak\i{}r}

\email{tcakir@mail.cern.ch}

\affiliation{Department of Physics, CERN, Geneva, Switzerland}

\author{S. Sultansoy}

\email{ssultansoy@etu.edu.tr}

\affiliation{Physics Division, TOBB University of Economics and Technology, Ankara,
Turkey }

\affiliation{Institute of Physics, Academy of Sciences, Baku, Azerbaijan}
\begin{abstract}
The single production of fourth family sneutrino $\tilde{\nu}_{4}$
via R-parity violating interactions in electron-positron collisions
has been investigated. We study the decays of $\tilde{\nu}_{4}$ into
different flavor dilepton $e^{\pm}\mu^{\mp}$ via R-parity violation.
It is shown that R-parity violating couplings $(\lambda_{411},\lambda_{412})$ down to 0.001 
will be reachable at future linear colliders which would provide better accuracy comparing 
to the indirect measurements as complementary to the LHC results.
\end{abstract}

\pacs{12.60.Jv Supersymmetric models\vspace{0.5cm}}

\maketitle
The addition of fourth family fermions (see recent reviews \cite{Holdom09}
and \cite{Sahin11}) to three families of the Standard Model (SM)
can be followed by the extension of minimal supersymmetric standard
model (MSSM) with the fourth family superpartners \cite{Carena96}
(we denote minimal supersymmetric standart model with three and four
families as MSSM3 and MSSM4, respectively). Concerning precision electroweak
data the parameter space of the MSSM4 is tightly constrained \cite{Murdock08,Godbole10}
if the neutrino has Dirac nature. However, this statement may be relaxed
if the neutrino has Majorana nature (as in the SM4 case \cite{Cobanoglu10},\cite{Asilar11}).

A search for supersymmetry (SUSY) is significant part of the physics
program of TeV scale colliders. As mentioned in \cite{Ari11}, it
is difficult to differentiate MSSM3 and MSSM4 at hadron colliders,
because the light superpartners of the third and fourth family quarks
has almost the same decay chains if the $R$-parity is conserved.
However, the rich phenomenology of the MSSM becomes even richer if
$R$-parity is violated (see \cite{Barbier05} and references therein).
The R-parity is defined as $R=(-1)^{3(B-L)-2s}$, where $B,\, L$
and $s$ are the baryon number, lepton number and spin, respectively.
Recently, searches for lepton flavor violating decays of third family
sneutrinos into different flavor dileptons have been performed by
the CDF and ATLAS experiments (see \cite{Aaltonen10} and \cite{ATLAS},
respectively). 

The baseline energy options for the linear colliders are assumed to
be $\sqrt{s}=0.5$ TeV for International Linear Collider (ILC) \cite{Brau07_v3}
and $\sqrt{s}=3$ TeV for Compact Linear Collider (CLIC) \cite{Assmann00}.
An energy option $\sqrt{s}=1$ TeV is also considered corresponding
either to the early stage of the CLIC or to an upgraded version of
the ILC. They have been designed to meet the requirements for planned
physics search programs \cite{Brau07_v2,Accomando04,CLIC_CDR11}.

In this work, we consider the process $e^{+}e^{-}\rightarrow e^{\pm}\mu^{\mp}$
for linear collider energies $\sqrt{s}=1$ TeV and $3$ TeV. There
are contributions to the cross section from both $s$-channel and
$t$-channel diagrams, which were not studied previously for the production
of fourth family sneutrino. The production cross section will depend
on the mixture of RPV couplings $\lambda_{411}$ and $\lambda_{412}$.

The R-parity violating extension of the MSSM superpotential is given
by

\begin{equation}
W_{RPV}=\frac{1}{2}\lambda_{ijk}\epsilon^{ab}L_{i}^{a}L_{j}^{b}\overline{E}_{k}+\lambda'_{ijk}\epsilon^{ab}L_{i}^{a}Q_{j}^{b}\overline{D}_{k}+\frac{1}{2}\lambda"{}_{ijk}\epsilon^{\alpha\beta\gamma}\overline{U}_{i}^{\alpha}\overline{D}_{j}^{\beta}\overline{D}_{k}^{\gamma}\label{eq:1}
\end{equation}
where $i,j,k=1,2,3,4$ are the family indices; $a,b=1,2$ are the
$SU(2)_{L}$ indices and $\alpha,\beta,\gamma$ are the $SU(3)_{C}$
indices. $L_{i}(Q_{i})$ are lepton (quark) $SU(2)$ doublet superfields;
$E_{i}(D_{i},U_{i})$ are the charged lepton (down-type and up-type
quark) $SU(2)$ singlet superfields. The couplings $\lambda_{ijk}$
and $\lambda"{}_{ijk}$ correspond to the lepton number violating
and baryon number violating interactions, respectively. Clearly, the
coupling constants $\lambda_{ijk}$ are antisymmetric under the exchange
of the first two indices, while the $\lambda"{}_{ijk}$ are antisymmetric
in last two indices. The first term in Eq. \ref{eq:1} leads to resonant
production of sneutrinos in lepton-lepton collisions, while the second
term allows slepton and sneutrino resonances in hadron-hadron collisions
\cite{Dreiner01}. The squark resonances can also be produced in the
lepton-hadron collisions \cite{Tang11}. Finally, the third term allows
resonant squark production in the hadron-hadron collisions. The magnitudes
of the RPV couplings are arbitrary, and they are restricted only from
the phenomenological considerations. A survey of the existing constraints
on the RPV couplings (for three families) can be found in Refs. \cite{Allanach99,Barbier05}.

The single production of fourth family sneutrino at the ILC and CLIC
proceeds via the interaction terms in the Lagrangian written in terms
of the component fields

\begin{equation}
L_{RPV}=-\lambda_{4jk}\tilde{\nu}_{4L}\ell_{jL}\bar{\ell}_{kR}+\lambda_{i4k}\tilde{\nu}_{4L}\ell_{iL}\bar{\ell}_{kR}+H.c.\label{eq:2}
\end{equation}
where $\tilde{\nu}_{4L}$ is the fourth family sneutrino field and
$\ell_{L(R)}$ is the left-handed (right-handed) lepton field, respectively.
Once produced in $e^{+}e^{-}$ collisions the fourth family sneutrino
$\tilde{\nu}_{4}$ can decay through different modes \cite{Dreiner01}:
RPV decays $\tilde{\nu}_{4}\to\ell_{j}^{+}\ell_{k}^{-}$ and $\tilde{\nu}_{4}\to\bar{d}_{j}d_{k}$,
supersymmetric decay $\tilde{\nu}_{4}\to\nu_{4}\tilde{\chi}^{0}$,
gauge decay $\tilde{\nu}_{4}\to\ell_{4}^{-}\tilde{\chi}^{+}$, weak
decay $\tilde{\nu}_{4}\to\tilde{\ell}_{4}^{-}W^{+}$ and Higgs decay
$\tilde{\nu}_{4}\to\tilde{\ell}_{4}^{-}H^{+}$.

Let us consider the signal from the decay $\tilde{\nu}_{4}\to e^{-}\mu^{+}$
with different flavor charged leptons, both of which are well isolated
and have high transverse momentum. The study is performed under the
hypothesis that only the fourth family sneutrino ($\tilde{\nu}_{4}$)
is produced and the sneutrino decay is determined by the $e{}^{\pm}\mu^{\mp}$
and $e^{+}e^{-}$ modes.

\begin{figure}
\includegraphics[scale=0.8]{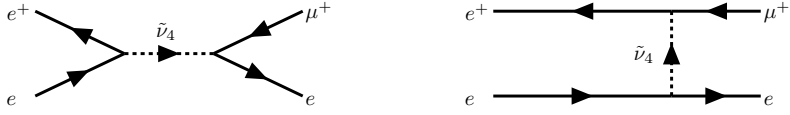} \caption{Feynman diagrams contributing to the process $e^{+}e^{-}\to e^{-}\mu^{+}$.
\label{fig:1}}
\end{figure}

The cross section for the process $e^{+}e^{-}\to e^{-}\mu^{+}$ ,
as shown in Fig.\ref{fig:1}, is given by

\begin{equation}
\sigma=\frac{(\lambda{}_{411}\lambda_{412}){}^{2}}{32\pi s^{2}}\left[\frac{s^{3}}{(s-m_{\tilde{\nu}_{4}}^{2})^{2}+m_{\tilde{\nu}_{4}}^{2}\Gamma_{\tilde{\nu_{4}}}^{2}}+\frac{s(s+2m_{\tilde{\nu}_{4}}^{2})}{s+m_{\tilde{\nu}_{4}}^{2}}+2m_{\tilde{\nu}_{4}}^{2}\log[\frac{m_{\tilde{\nu}_{4}}^{2}}{s+m_{\tilde{\nu}_{4}}^{2}}]\right]\label{eq:8}
\end{equation}
 where $m_{\tilde{\nu}_{4}}$ and $\Gamma_{\tilde{\nu}_{4}}$ are
the mass and decay width of fourth family sneutrino, respectively.
We calculate the decay width of fourth family sneutrino depending
on its mass and the RPV couplings with the assumption of the relevant
coupling dominance

\[
\Gamma_{\tilde{\nu}_{4}}=(\lambda_{411}^{2}+2\lambda_{412}^{2})m_{\tilde{\nu}_{4}}/16\pi
\]

\begin{figure}
\includegraphics[scale=0.8]{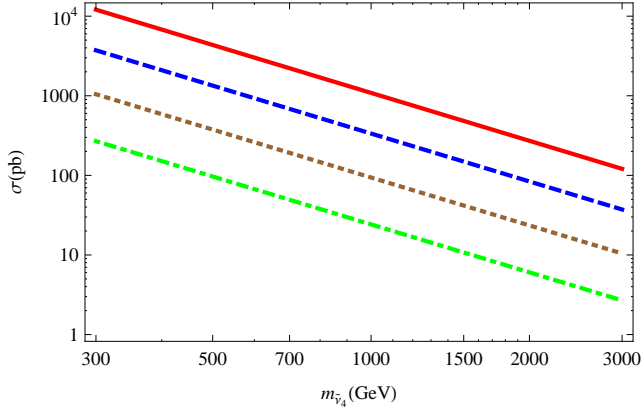}

\caption{The resonance production cross section for the fourth family sneutrino
with RPV couplings $\lambda_{412}=\lambda_{411}$(solid, red), $\lambda_{412}=0.01,\lambda_{411}=0.05$
(dashed, blue), $\lambda_{412}=0.05,\lambda_{411}=0.01$ (dotted,
brown) and $\lambda_{412}=0.01,\lambda_{411}=0.001$ (dotdashed, green)
at a linear collider. \label{fig:fig2}}
\end{figure}

Assuming the couplings $\lambda_{412}=0.1$($0.05$), $\lambda_{411}=0.1$($0.01$)
and the mass $m_{\tilde{\nu}_{4}}=1$ TeV we calculate the decay width
$\Gamma_{\tilde{\nu}_{4}}=0.597$ ($0.101$) GeV. It can also be scaled
for other mass values. The cross section for resonance production
of the fourth family sneutrino for different RPV couplings is shown
in Fig. \ref{fig:fig2}. 

For numerical calculations we implement the vertices from interaction
Lagrangian (Eq. 2) into CalcHEP \cite{CalcHEP}, and we take into
account the effects from initial state radiation (ISR) and beamstrahlung
(BS) using the beam parameters as shown in Table \ref{tab:table1}.

\begin{table}
\caption{The collider beam parameters of the ILC and CLIC needed to calculate
the ISR+BS effects. \label{tab:table1}}

\begin{tabular}{|c|c|c|}
\hline 
 & ILC & CLIC\tabularnewline
\hline 
\hline 
Center of mass energy (TeV) & $1$ & $3$\tabularnewline
\hline 
Horizontal beam size (nm) & $640$ & $45$\tabularnewline
\hline 
Vertical beam size (nm) & $5.7$ & $1$\tabularnewline
\hline 
Bunch length (mm) & $0.3$ & $0.044$\tabularnewline
\hline 
Number of particles in the bunch (N) & $2\times10^{10}$ & $3.72\times10^{9}$\tabularnewline
\hline 
Design luminosity (cm$^{-2}$s$^{-1}$) & $2\times10^{34}$ & $5.9\times10^{34}$\tabularnewline
\hline 
\end{tabular}
\end{table}

By fixing the RPV couplings $\lambda_{412}=\lambda_{411}=0.1$, the
signal cross sections for the fourth family sneutrino production for
different mass values are presented in Table \ref{tab:table2}, 
where the cross sections $\sigma_{\mbox{ISR+BS}}$
include initial state radiation (ISR) and beamstrahlung (BS) in $e^{+}e^{-}$
collisions at $\sqrt{s}=1$ TeV and $\sqrt{s}=3$ TeV. For smaller
RPV couplings such as $\lambda_{412}=\lambda_{411}=0.01$, assuming
$m_{\tilde{\nu_{4}}}=400$ GeV we calculate the signal cross sections
as $4.30\times10^{-3}$ pb and $1.38\times10^{-2}$ pb for $\sqrt{s}=1$
and $3$ TeV, respectively.

\begin{table}
\caption{The signal cross section for the process $e^+e^-\to e^-\mu^+$ via fourth family sneutrino exchange. 
Here, we assume RPV couplings $\lambda_{412}=\lambda_{411}=0.1$.
The cross sections $\sigma_{ISR+BS}$ include initial state radiation
(ISR) and beamstrahlung (BS), in $e^{+}e^{-}$ collisions at $\sqrt{s}=1$
TeV and 3 TeV.\label{tab:table2}}

\begin{tabular}{|c|c|c|}
\hline 
 & $\sqrt{s}=1$ (TeV) & $\sqrt{s}=3$ (TeV)\tabularnewline
Mass (GeV) & $\sigma_{\mbox{ISR+BS}}$(pb) & $\sigma_{\mbox{ISR+BS}}$(pb)\tabularnewline
\hline 
\hline 
200 & $1.70\times10^{-1}$ & $2.23\times10^{0}$\tabularnewline
\hline 
400 & $1.93\times10^{-1}$ & $1.42\times10^{0}$\tabularnewline
\hline 
600 & $2.82\times10^{-1}$ & $1.11\times10^{0}$\tabularnewline
\hline 
800 & $8.58\times10^{-1}$ & $7.69\times10^{-1}$\tabularnewline
\hline 
1000 & $1.70\times10^{2}$ & $4.96\times10^{-1}$\tabularnewline
\hline 
1200 & $1.37\times10^{-3}$ & $3.28\times10^{-1}$\tabularnewline
\hline 
1400 & $3.34\times10^{-4}$ & $2.22\times10^{-1}$\tabularnewline
\hline 
1600 & $1.36\times10^{-4}$ & $1.56\times10^{-1}$\tabularnewline
\hline 
1800 & $7.00\times10^{-5}$ & $1.14\times10^{-1}$\tabularnewline
\hline 
2000 & $4.10\times10^{-5}$ & $8.81\times10^{-2}$\tabularnewline
\hline 
2200 & --- & $7.13\times10^{-2}$\tabularnewline
\hline 
2400 & --- & $6.15\times10^{-2}$\tabularnewline
\hline 
2600 & --- & $5.82\times10^{-2}$\tabularnewline
\hline 
2800 & --- & $6.69\times10^{-2}$\tabularnewline
\hline 
3000 & --- & $4.06\times10^{0}$\tabularnewline
\hline 
\end{tabular}
\end{table}

\begin{figure}
\includegraphics[scale=0.625]{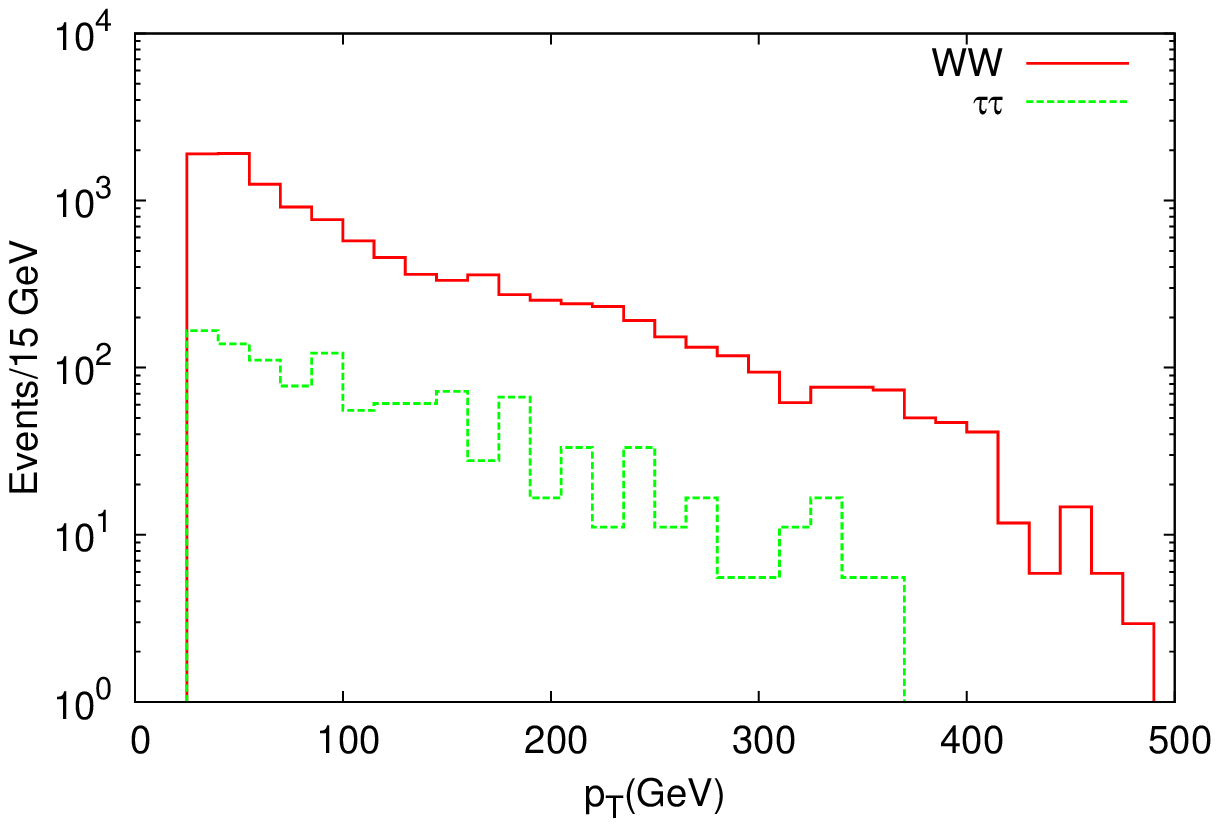}
\includegraphics[scale=0.625]{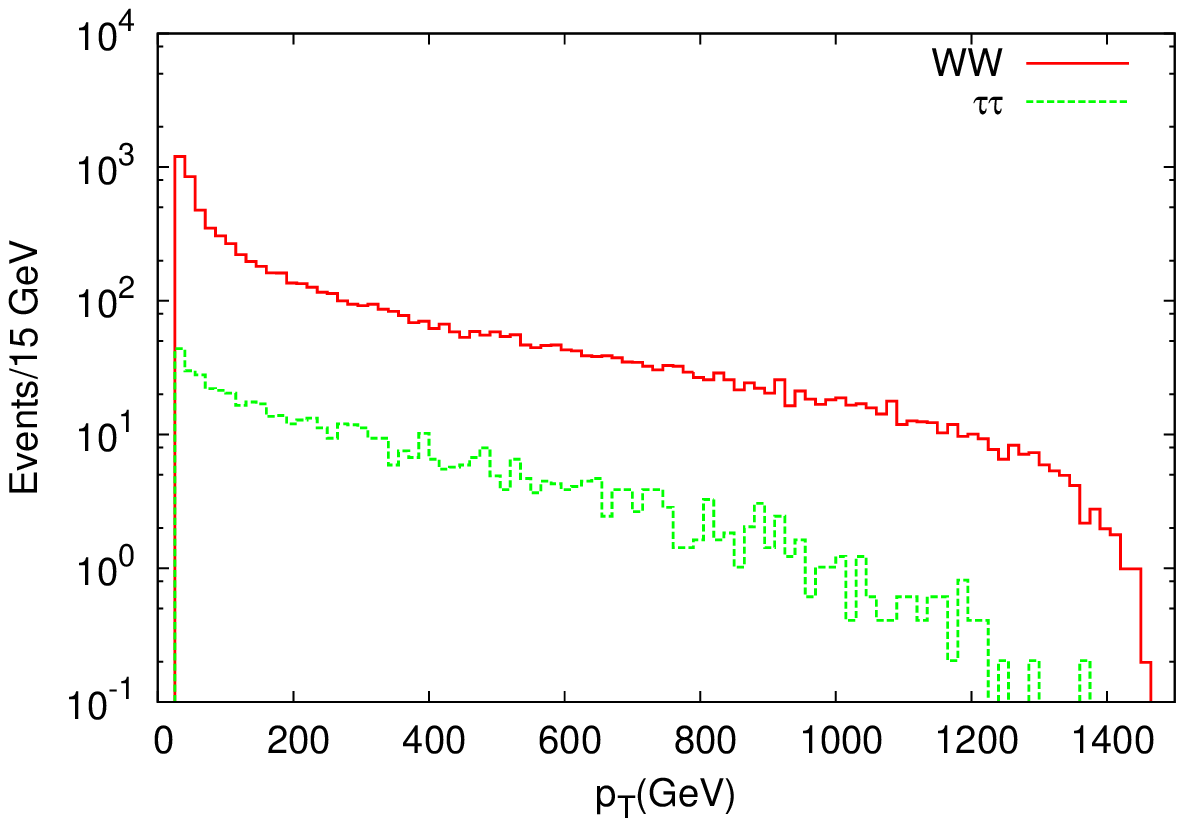}
\caption{The transverse momentum distribution for the background (with ISR effects) at 
$\sqrt{s}=1$ TeV (left) and $\sqrt{s}=3$ TeV (right). \label{fig:fig3}}
\end{figure}

The main contributions to the background comes from the pair production
of $W^{+}W^{-}$ and $\tau^{+}\tau^{-}$. The cross section for top-pair
production $1.73\times10^{-1}$($2.03\times10^{-1}$) pb at $\sqrt{s}=1$
TeV and $1.98\times10^{-2}$($1.81\times10^{-1}$) pb at $\sqrt{s}=3$
TeV without (with) ISR+BS effects, respectively. This background is
an order smaller than the $W^{+}W^{-}$ background and it can be removed
by veto on high energy jets from two $b$-quarks. The cross sections
for the $W^{+}W^{-}$ and $\tau^{+}\tau^{-}$ backgrounds as calculated
with PYTHIA \cite{PYTHIA} including the ISR effects are given in
Table \ref{tab:table3} for the center of mass energies $\sqrt{s}=1$
and $3$ TeV. The transverse momentum distributions of charged leptons ($e^-$ or $\mu^+$) 
in the final state for $WW$ and $\tau\tau$ backgrounds are shown in 
Figure \ref{fig:fig3} for the ILC and CLIC energies.

\begin{table}
\caption{Calculated backgrounds for the pair production of $W^{+}W^{-}$and
$\tau^{+}\tau^{-}$ at linear colliders. The numbers correspond to the cross sections 
with the ISR effects. \label{tab:table3}}

\begin{tabular}{|c|c|c|}
\hline 
$\sqrt{s}$(TeV) & $\sigma(W^{+}W^{-})$(pb) & $\sigma(\tau^{+}\tau^{-})$(pb)\tabularnewline
\hline 
1 & $3.17\times10^{0}$ & $3.36\times10^{-1}$\tabularnewline
\hline 
3 & $4.07\times10^{0}$ & $1.29\times10^{0}$\tabularnewline
\hline 
\end{tabular}
\end{table}

The contributions from these backgrounds to the final state $e^{\pm}\mu^{\mp}+X$
are estimated in the invariant mass distribution. In order to make
the analysis with the signal and background it is required that electron
and muon have transverse momentum $p_{T}^{e,\mu}>25$ GeV and pseudorapidity
$|\eta_{e,\mu}|<2.5$. For these cuts the signal cross section reduces at most $6\%$, while 
the WW background reduces by $25\%$ and $\tau\tau$ background reduces by $40\%$.

\begin{figure}
\includegraphics{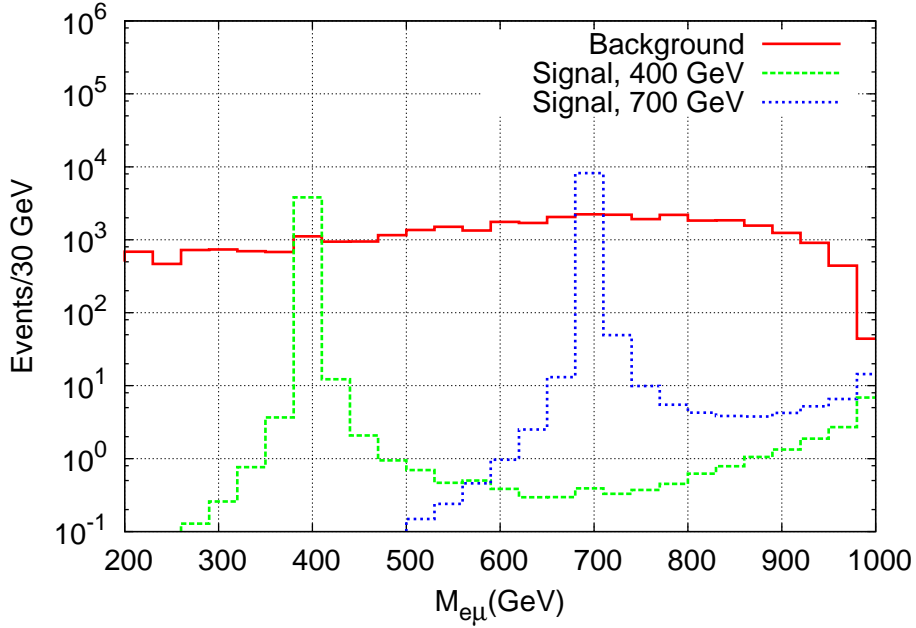}

\caption{The $e\mu$ invariant mass distribution for background (with ISR effects) and signal (with ISR and BS effects)
at $\sqrt{s}=1$ TeV. \label{fig:fig4}}
\end{figure}

\begin{figure}
\includegraphics{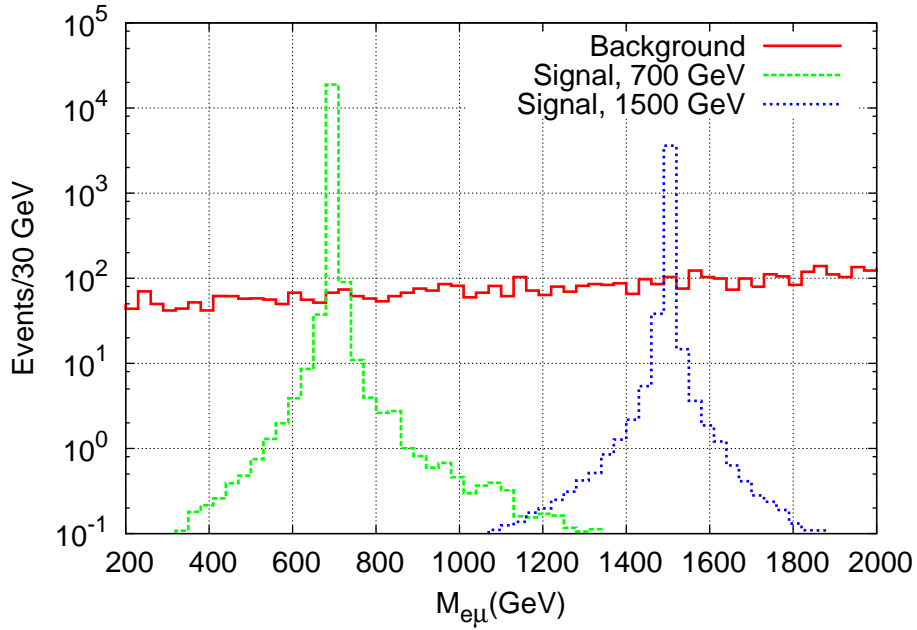}

\caption{The $e\mu$ invariant mass distribution for background (with ISR effects) and signal (with ISR and BS effects) 
at $\sqrt{s}=3$ TeV. \label{fig:fig5}}
\end{figure}

The invariant mass distributions of $e\mu$ system are shown in Fig.
\ref{fig:fig4} at $\sqrt{s}=1$ TeV and in Fig. \ref{fig:fig5} at
$3$ TeV for both background and signal (assuming $\lambda_{411}=\lambda_{412}=0.1$).
From these figures we see that background remains almost at the same
level for interested mass region. Therefore, we examined the invariant
mass intervals for the existence of a fourth family sneutrino signal.
The search region in the invariant mass spectrum is divided into mass
bins $\Delta m$ for specific $m_{\tilde{\nu}_{4}}$, which is defined
to be ($m_{\tilde{\nu_{4}}}\pm3\sigma$), where $\sigma$ is the mass
resolution.

In Tables \ref{tab:table4} and \ref{tab:table5}, we present the
cross sections $\Delta\sigma$ for signal and background within the
mass intervals centered at each mass values. The statistical significance
$SS=S/\sqrt{B}$ for signal observation is presented in the last column
of these tables. 

\begin{table}
\caption{The cross sections for the signal and background calculated within
the mass intervals at the center of mass energy $\sqrt{s}=1$ TeV.
Here, we assume the RPV couplings $\lambda_{411}=\lambda_{412}=0.005$.
The statistical significance is calculated for an integrated luminosity
200 fb$^{-1}$. \label{tab:table4}}

\begin{tabular}{|c|c|c|c|c|}
\hline 
$m_{\tilde{\nu}_{4}}$(GeV) & $\Delta m$(GeV) & \multicolumn{1}{c|}{$\Delta\sigma_{s}$(fb)} & $\Delta\sigma_{B}$(fb) & \multicolumn{1}{c|}{$SS$}\tabularnewline
\hline 
\hline 
200 & 20 & $4.22\times10^{-1}$ & $8.34\times10^{-1}$ & $6.5$\tabularnewline
\hline 
400 & 30 & $4.82\times10^{-1}$ & $1.60\times10^{0}$ & $5.4$\tabularnewline
\hline 
600 & 40 & $6.99\times10^{-1}$ & $2.23\times10^{0}$ & $6.6$\tabularnewline
\hline 
800 & 50 & $2.14\times10^{0}$ & $3.56\times10^{0}$ & $16.1$\tabularnewline
\hline 
1000 & 60 & $6.57\times10^{4}$ & $7.35\times10^{-2}$ & $5.6\times10^{3}$\tabularnewline
\hline 
\end{tabular}
\end{table}

\begin{table}
\caption{The same as Table \ref{tab:table4}, but for $\sqrt{s}=3$ TeV and
$L_{int}=600$ fb$^{-1}$. \label{tab:table5}}

\begin{tabular}{|c|c|c|c|c|}
\hline 
$m_{\tilde{\nu}_{4}}$(GeV) & $\Delta m$(GeV) & \multicolumn{1}{c|}{$\Delta\sigma_{s}$(fb)} & $\Delta\sigma_{B}$(fb) & \multicolumn{1}{c|}{$SS$}\tabularnewline
\hline 
\hline 
200 & 20 & $5.73\times10^{0}$ & $2.79\times10^{0}$ & $84.0$\tabularnewline
\hline 
400 & 30 & $3.60\times10^{0}$ & $2.32\times10^{0}$ & $57.9$\tabularnewline
\hline 
600 & 40 & $2.79\times10^{0}$ & $2.77\times10^{0}$ & $41.1$\tabularnewline
\hline 
800 & 50 & $1.92\times10^{0}$ & $3.20\times10^{0}$ & $26.3$\tabularnewline
\hline 
1000 & 60 & $1.24\times10^{0}$ & $4.50\times10^{0}$ & $14.3$\tabularnewline
\hline 
1200 & 70 & $8.30\times10^{-1}$ & $3.95\times10^{0}$ & $10.2$\tabularnewline
\hline 
1400 & 80 & $5.50\times10^{-1}$ & $4.82\times10^{0}$ & $6.1$\tabularnewline
\hline 
1600 & 90 & $3.80\times10^{-1}$ & $5.71\times10^{0}$ & $3.9$\tabularnewline
\hline 
1800 & 100 & $2.90\times10^{-1}$ & $5.81\times10^{0}$ & $2.9$\tabularnewline
\hline 
2000 & 110 & $2.10\times10^{-1}$ & $5.81\times10^{0}$ & $2.1$\tabularnewline
\hline 
2200 & 120 & $1.70\times10^{-1}$ & $6.68\times10^{0}$ & $1.6$\tabularnewline
\hline 
2400 & 130 & $1.50\times10^{-1}$ & $7.12\times10^{0}$ & $1.4$\tabularnewline
\hline 
2600 & 140 & $1.40\times10^{-1}$ & $4.71\times10^{0}$ & $1.6$\tabularnewline
\hline 
2800 & 150 & $1.60\times10^{-1}$ & $2.96\times10^{0}$ & $2.3$\tabularnewline
\hline 
3000 & 160 & $1.61\times10^{3}$ & $5.50\times10^{-1}$ & $5.3\times10^{4}$\tabularnewline
\hline 
\end{tabular}
\end{table}

\begin{figure}
\includegraphics{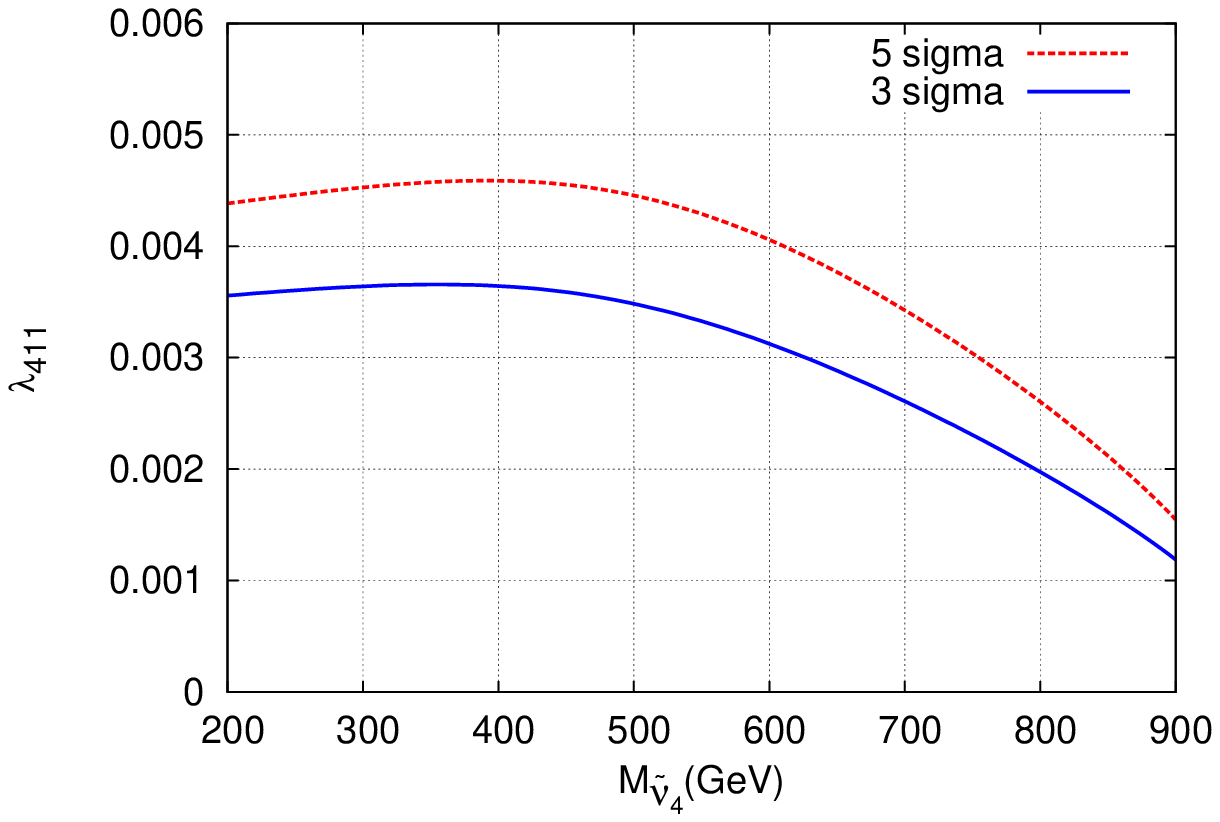}

\caption{Attainable limits for the mass and RPV couplings of fourth family
sneutrino at $\sqrt{s}=1$ TeV and $L_{int}=200$ fb$^{-1}$ (assuming $\lambda_{412}=\lambda_{411}$).
\label{fig:fig6}}
\end{figure}

\begin{figure}
\includegraphics{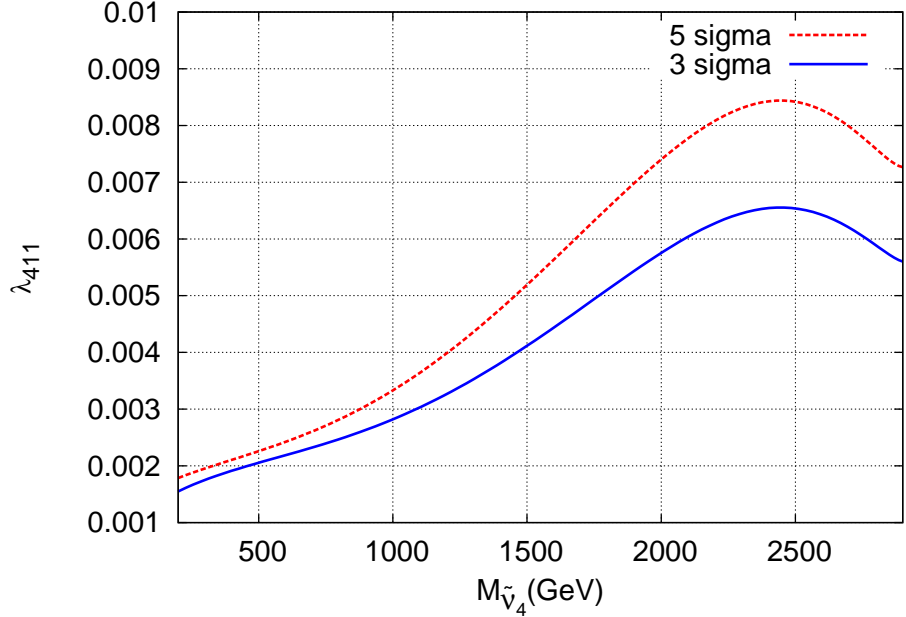}

\caption{Attainable limits for the mass and RPV couplings of fourth family
sneutrino at $\sqrt{s}=3$ TeV and $L_{int}=600$ fb$^{-1}$ (assuming $\lambda_{412}=\lambda_{411}$).
\label{fig:fig7}}
\end{figure}

In Figs. \ref{fig:fig6} and \ref{fig:fig7}, we present contour plots
in the ($m_{\tilde{\nu}_{4}}-\lambda_{411}$) plane for an integrated
luminosity of $200$ fb$^{-1}$ at $\sqrt{s}=1$ TeV and for an integrated
luminosity $600$ fb$^{-1}$ at $\sqrt{s}=3$ TeV, respectively. The
regions above the curve denotes the range of RPV coupling $\lambda_{411}$
values that can be reached in the linear collider experiments. It
is seen that even for $m_{\tilde{\nu}_{4}}<\sqrt{s}$ values the RPV
couplings well below 0.01 are reachable.

In order to compare the potential of ILC and CLIC, we assume the mass 
value $m_{\tilde{\nu}_4}=500$ GeV and the RPV couplings $\lambda_{412}=\lambda_{411}$. 
In this case, one year of operation will give opportunity to 
reach $\lambda_{411}=0.0035$ at ILC and $0.002$ at CLIC. However, 
in order to reach the same sensitivity as CLIC, the ILC should operate 9 years.

In conclusion, the resonance production of fourth family sneutrino
through R-parity violating couplings at the linear collider energies
have been studied. The sensitivity to RPV couplings can be measured
with better level compared to the information derived from indirect
measurements. The results show that RPV couplings of fourth family
sneutrino in the high mass region can be explored at linear collider
experiments as complementary to the LHC results.

\begin{acknowledgments}
This work is partially supported by Turkish Atomic Energy Authority
(TAEK) and State Planning Organization (DPT). 
\end{acknowledgments}

\end{document}